# Application of Percolation Theory to Current Transfer in Granular Superconductors

B. Zeimetz[a], B. A. Glowacki, and J. E. Evetts

Department of Materials Science and IRC in Superconductivity, Cambridge University, Pembroke Street, Cambridge CB2 3QZ, U. K.
[a] e-mail bpz20@cam.ac.uk, phone +44-1223-334375, fax +44-1223-334373

**Short Title:**
Percolation Theory and Current Transfer in Granular Superconductors

## Abstract

We investigate the description of current transfer in polycrystalline superconductors by percolation theory and its limitations. Various computer models that have been proposed are reviewed and related to the experimental and theoretical framework. While some conductor properties can be well described by percolation theory and models, we argue that a conceptual gap exists between experiment and theory. This gap has to be bridged by finding relations between electromagnetic and statistical parameters. We derive various such relations and compare them with recent simulation data. In particular, we suggest a new scaling law between the two fundamental variables 'current' and 'probability'.

## 1 Introduction

The percolation concept was first employed to describe superconductors by Davidson and Tinkham [1], who analysed resistivity data of composite $Nb_3Sn/Cu$ wires. Deutscher *et al.* [2] studied temperature dependent properties of granular Al films with $Al_2O_3$ grain boundaries (GBs). The superconducting transition of individual GBs is then controlled by thermal fluctuations. Percolation theory has also been used in a number of studies to analyse mixtures of superconductors and non-superconductors, prepared by powder mixing and sintering [3] or inhomogeneous doping [4]. A transition can also be induced by variation of magnetic field [5][6] or film thickness [6]. The movement of Abrikosov vortices, which are subjected to an inhomogeneous local pinning force, has also been described as a percolative process [7,8]. In all these cases, the state of an individual grain grain boundary or phase boundary is controlled by random fluctuations independently of its neighbours, which is the basic assumption of percolation theory.

However, the situation is different when the superconductor-to-normal transition is induced by a transport current [9,10]: now the Kirchhoff rules have to be considered, and the probability of individual GBs becoming non-superconducting is no longer independent. One expects correlation effects, and percolation theory cannot be applied straightforwardly. However, the behaviour of a polycrystalline superconductor near its critical current (at constant temperature and external field) is the most interesting case from a practical point of view, because of its relevance to application of (granular) high temperature superconductors (HTS) as conductors in power transmission, magnets etc.

In this article, we discuss the compatibility between percolation theory and current transfer near and above the critical current. We focus mainly on the case where GBs are responsible for percolative current flow.

## 2 Review of Percolation Concepts in Granular Superconductors

### 2.1 Percolation Theory and Random Resistor Networks

Percolation theory [11-13] deals with $D$-dimensional networks of 'sites', each of which is connected to its neighbours by 'bonds'. If the network forms a regular lattice, which is the case most frequently considered, the number of bonds for each site is the coordination number $Z$ of the lattice. This paper deals only with 'bond percolation', where one of the properties 'occupied' or 'empty' is attributed to each bond. The fundamental variable of percolation theory is the probability $p$. Each bond is occupied with probability $p$ and empty with probability $1-p$. The states are set *randomly and independently* of each other. Two neighbouring sites are 'connected' if they share an occupied bond, and groups of connected sites form 'clusters'. Above a critical probability $p_c$, which for an infinite network depends only on the geometry and dimension, a single 'infinite cluster' spans the entire system. This picture leads to a general desciption of critical points and phase transitions, mainly concerning scaling laws near the critical point. For bond percolation, the product $Z \times p_c$ is approximately constant: $Z \times p_c \approx 2$ in $2D$, and $Z \times p_c \approx 1.5$ in $3D$.

While all the above is based purely on geometrical arguments, it can be applied to electrical conduction in disordered media using the concept of random resistor networks [14,15]. In a random resistor network, the bonds are replaced by resistors, and a resistance $R_B$ is assigned *randomly* to each of them. In the case of mixed superconductor/normal conductor networks, the resistance is either $R_B = 0$ (with probability $p$) or $R_B = R_0 > 0$ (with probability 1–$p$). Hence in such a network one finds superconducting and normal conducting clusters of resistors, and the average cluster size is determined by $p$.

In analogy to geometric percolation, a critical probability $p_c$ separates two domains of the global transport properties: The size of the superconducting clusters is characterised by the correlation length $\xi$, which scales as $\xi \sim |p{-}p_c|^{-\nu}$ for $p < p_c$. For $p > p_c$, the network can carry a global transport supercurrent. If a finite critical current $I_{cB}$ is assigned to each of the superconducting resistors, the overall network also has a finite, global critical current density $J_{cN}$, which scales as [11-13]

$$J_{cN}(p) \sim (p{-}p_c)^{\nu} \qquad \text{for } p >\approx p_c, \qquad (1)$$

i.e. with the same exponent as the correlation length. (Throughout this paper we use the index B to denote local properties of individual bonds or grain boundaries, and the index N for global network properties) Below $p_c$, the system has a finite resistivity $\rho$, which near $p_c$ scales as

$$\rho(p) \sim |p{-}p_c|^{s} \qquad \text{for } p <\approx p_c \qquad (2)$$

The scaling exponents $\nu$ and $s$ are universal and depend only on the dimensionality of the network. For $D = 2$, $\nu = 4/3$ and $s \approx 1.30$, while for $D = 3$, $\nu \approx 0.875$ and $s \approx 0.73$.

If the network has a finite size, all the above properties are modified due to fluctuation effects. For regular lattices and equal dimensions, the size can be described by the number of elements $N$ along each dimension, and the total number of elements is $N^D$. The resistivity then scales with the size as $\rho \sim N^{s/\nu}$, and $p_c$ is reduced by

$$p_c(\infty) - p_c(N) \sim N^{-1/\nu} \qquad (3)$$

($p_c(N)$ has to be read as the average value taken over many samples of size $N$).

The concept of correlated percolation [16-18] has been introduced to study percolative systems with interaction between the lattice sites. It is mainly used in connection with the Ising model to study e.g. diluted ferromagnets and sol-gel transitions. In these cases the interaction is restricted to nearest neighbours, and can be expressed as a constant parameter (e.g. spin exchange energy $J$). One can then identify phase transitions as function of temperature and concentration, and scaling laws similar to those of 'random percolation'.

## 2.2 Superconducting Grain Boundaries and Granular Superconductors

In many superconductors, especially high-temperature superconductors (HTS), the superconducting properties are considerably weakened at GBs compared to intra-grain values [19-21]. Namely the critical current density $J_c$ is strongly reduced at high GB angles $\alpha$, typically following an exponential decrease with a plateau at low angles:

$$J_{cB}(\alpha) = J_{c0} \qquad\qquad\qquad \text{for } \alpha \leq \alpha_c, \qquad (4a)$$

$$J_{cB}(\alpha) = J_{c0} \times \exp(-(\alpha{-}\alpha_c)/\alpha_0) \qquad \text{for } \alpha \geq \alpha_c. \qquad (4b)$$

This dependence is well-established for $YBa_2Cu_3O_7$ (YBCO) [21][22], and recently has also been confirmed in $Bi_2Sr_2Ca_2Cu_3O_{10}$ (Bi2223) [23]. The values for the critical angles $\alpha_c$ and $\alpha_0$ are around $5^o$ but vary considerably between different materials and processes [21-25]. For the discussion to follow, we note that for $\alpha_0 \rightarrow 0$

$$J_{cB}(\alpha) = J_{c0} \qquad \text{for } \alpha \leq \alpha_c, \tag{5a}$$

$$J_{cB}(\alpha) = 0 \qquad \text{for } \alpha > \alpha_c, \tag{5b}$$

In YBCO thin film low-angle grain boundaries ($\alpha \sim 5^o\text{-}10^o$) subjected to magnetic fields ($B \sim 1$ T), the dissipation at GB is attributed to one-dimensional viscous flux flow [26]. The GB is penetrated by a linear row of flux vortices, which are oriented perpendicular to the film. This mechanism leads to a linear current-voltage (*I-V*) characteristics of the GB:

$$V(I > I_{cB}) = R_{ff} (I - I_{cB}), \tag{6}$$

where $R_{ff}$ is the flux flow resistance and $I_{cGB}$ the critical current of the GB. More recently, Hogg *et al.* [27] showed that the *I-V* curve can also exhibit a number of kinks, and explained this by depinning of several parallel rows of flux lines.

All macroscopic HTS conductors consist of polycrystalline material, and therefore form networks of GBs [28][29]. The importance of GBs for current transfer and dissipation depends both on the superconducting material and on the processing of the conductors. Generally the processing is aimed at achieving high texture, in order to minimise the GB angles, cf. Eq. 4.

YBCO coated conductors consist of a polycrystalline, textured film on a metal substrate, i.e. they essentially are 2-*D* conductors [30]. By detailed electromagnetic and magneto-optical investigations it was established that, in coated conductors the dissipation is dominated by GBs [28-32]. Magneto-optical images show that the GBs are preferentially penetrated by magnetic flux, Fig. 1, indicating that during current transfer the GBs form a channel network for flux lines, and the same mechanism as in low angle GBs is at work [28].

In BiSCCO powder-in-tube tapes [33] and other Bi-based HTS conductors, the importance of GBs is much more contentious [10]. A powder-in-tube tape is a 3-*D* grain network, which makes direct local magneto-optical observations [34] much more difficult. It is furthermore clear that in these materials, intra-grain critical currents are low in an applied magnetic field, and might dominate dissipation [35].

We will discuss some of our experimental data for both types of conductors in Section 4.

## 2.3 Network Models for Granular Superconductors

A number of network models for computer simulation have been proposed to investigate properties of granular HTS [36-46]. Most of these models deal with regular lattices, such as squares, cubes, hexagons etc. The grain orientations and/or GB angles are set arbitrarily, e.g. using a random number generator to create a Gaussian distribution. In the case of 2-*D* coated conductors it is also possible to directly measure the local texture and thereby construct an experimental grain map including grain orientations. Some computer models use these experimental data as input for calculations [43,45,46].

In terms of the model output, one can distinguish between work which only determines the onset critical current of a network, and a few models which construct a nonlinear resistor network to calculate the entire *I-V* curve [38,45,46].

In the simple 'bimodal' case, the GBs are assigned the values 'superconducting' or 'resistive', depending on their GB angle relative to a threshold value $\alpha_c$. The bimodal model can be summarised by Eq. 5. If the GB angles are independently and randomly set (see below), this case also represents a conventional random resistor network as described above. The onset critical current $I_{c0}$ of the conductor, at which dissipation sets in, is then determined by the 'limiting cross section' [40], which is called the 'minimum cut' in network theory.

In the same spirit, experimentally measured grain maps are sometimes drawn as 'percolation maps' for different values of $\alpha_c$ [41,47]. Areas connected by GBs with $\alpha < \alpha_c$ are typically shown in single color, and blocking GBs as black lines. An example is shown in Fig. 2, with the largest 'connected' area/cluster marked grey, and the 'limiting cross section' indicated with arrows. Fig. 2 demonstrates that typically the ratio of connected area to total area is much larger than the limiting cross section relative to the conductor width. In percolation theory, this fact is described by distinguishing between the 'cluster mass' and the 'cluster backbone mass' [11,12]. The latter excludes the 'dead ends' of a cluster, which do not contribute to conduction. Due to the 'fractal geometry' of the clusters, the transport properties of percolative systems are very different from those of ordinary lattices.

This simplified model has been supported by magneto-optical measurements on YBCO coated conductors by Feldmann et al [47]. They found that for a given field and temperature, only GBs with angles $\alpha$ above a sharply defined critical angle $\alpha_{c,exp}$ were penetrated by magnetic flux ($\alpha_{c,exp} = 4^o$ at 60 mT and 15 K). The setup of this measurement did not include a transport current, and furthermore, magneto-optics shows flux density rather than flux flow. However, recent results have shown that, magneto-optical images using external fields and transport currents coincide surprisingly well with each other [31].

A more realistic model can be constructed by assigning each GB a *finite* critical current using equations (4) instead of (5). As before, the critical current $I_{c0}$ can be found by determining the 'minimum cut', for example by using a 'mouse walk' algorithm [43]. We note that this network is already different from the random resistor network described in section 2.1, since now *every* GB/bond can carry a supercurrent.

If one wants to determine the entire *I-V* characteristic of a network, each GB has to be described in terms of its nonlinear *I-V* curve. Haslinger and Joynt [38] have described a nonlinear resistor network where the GBs can have either the *I-V* curve described in Eq. (6), or a Josephson-junction type *I-V*.

We have proposed a model [45,46] where each GB is described by a pair of linear resistors, one with zero resistance but finite current capacity (defined by the GB critical current), and the second having resistance proportional to the flux flow resistivity. (This approach was adopted from network analysis, where it is used to approximate nonlinear transport properties [48,49]. In our case, it exactly reproduces Eq (6)). Some results of this model will be shown below.

We have so far described network models for superconductors only in terms of electromagnetic parameters ($\alpha_0$, $\alpha_c$, $I_c$, *I-V*). Some qualitative connections to percolation theory have already been made, but for a quantitative description we have to relate the electromagnetic parameters to the probability $p$ which is the fundamental variable of percolation theory. This will be attempted in the next section.

## 3 Relations between Electromagnetic and Statistical Parameters

We first need to define the probability $p$ more precisely. In percolation theory, $p$ is the independent variable of the system, and quantities are expressed as function of $p$. For example, in a random resistor network, $p$ is the probability that a resistor/GB is in the superconducting state. It is assumed that the states of the individual resistors/GBs are staistically independent.

In this section, however, we want to express $p$ as a function of other variables, namely, the electromagnetic parameters of a conductor. (In the terminology of statistics, $p$ is then the integrated probability distribution function) Furthermore, we investigate the statistical independence of the individual GB's state. We define $p$ *globally* as the number of superconducting GBs divided by the total number of GBs in a conductor. This definition is meaningful even if the individual states of the GBs are not independent.

### 3.1 Probability vs. Critical Angle $\alpha_c$

In the case of a bimodal model (Eq. 5, $\alpha_0 = 0$), the state of the GBs is determined by the GB angle and the critical angle $\alpha_c$. A GB is superconducting if the GB angle $\alpha \leq \alpha_c$, and otherwise the GB is normal conducting. Hence in this case the states of the individual GBs are statistically independent from each other, if the GB angles are statistically independent. $p(\alpha_c)$ is then equal to the probability that a randomly chosen GB has an angle smaller than $\alpha_c$.

In experimental conductors, the GB angles are independent insofar as the individual grain orientations are determined by fluctuations in the production process. (The question of correlation between neighbouring grains was discussed by Rutter et al. [41,50]) In a model structure for computer simulations, one can generate statistically independent GB angles by using a random number generator.

For the bimodal model and a known GB angle distribution, the probability distribution function $p(\alpha_c)$ can be calculated directly. From this distribution, we can calculate the onset critical current $I_{cN}(\alpha_c)$ of a conductor. In coated conductors, the GB angle is usually characterised as function of 'in-plane' and 'out-of plane' misalignment angles $\varphi$ and $\phi$, which refer to the grain misalignment relative to the conductor surface. If these two alignment angles have Gaussian distributions [41], and the GB angle is approximated by $\alpha = (\varphi^2 + \phi^2)^{1/2}$, its distribution is given by

$$F(\alpha) \sim \alpha/\alpha_m \times \exp(-\tfrac{1}{2}(\alpha/\alpha_m)^2) \qquad (7)$$

The maximum of the $F(\alpha)$ distribution is at $\alpha_m$, which is a function of the full half width of the misalignment angles. The distribution $F(\alpha,\alpha_m)$ is known as a Rayleigh distribution, and shown in the inset to Fig. 3, together with a distribution measured on a typical coated conductor. The probability of an angle being smaller than $\alpha_c$ is then obtained by integrating from 0 to $\alpha_c$:

$$p(\alpha<\alpha_c) \sim 1 - \exp(-\tfrac{1}{2}(\alpha_c/\alpha_m)^2) \qquad (8)$$

This function is shown in the main plot of Fig. 3. We can now apply percolation theory to find the limiting threshold angle $\alpha_{c0} = \alpha_c(p=p_c)$ above which the onset critical current $I_{cN}$ is nonzero, recalling that for $p < p_c$, there is no sample spanning cluster, i.e. no global supercurrent. If we model the coated conductors as a regular set of squares ($Z = 4$; $p_c = 1/2$) or hexagons ($Z = 6$; $p_c = 1/3$), we find from Fig. 3 that $\alpha_{c0}/\alpha_m = 1.17$ for squares and $\alpha_{c0}/\alpha_m = 0.9$ for hexagons. We conclude that, as a rule of thumb,

$$\alpha_{c0} \approx \alpha_m \qquad \text{or} \qquad p(\alpha_c = \alpha_m) \approx p_c \qquad (9)$$

i.e. for $\alpha_c < \alpha_m$ the transport critical current is zero in the bimodal model. In the real conductor of Fig. 2, with varying number of GB per grains, this result is confirmed: Here $\alpha_m \approx 4.5°$, and the supercurrent can percolate for $\alpha_c \geq 5$, hence $\alpha_{c0}/\alpha_m \approx 1.1$. Similar studies by Rutter et al [41, 50] on a hexagonal model also confirm our general rule, see Fig. 4.

For the critical current density dependence above $p_c$, percolation theory predicts $J_{cN}(p) \sim (p-p_c)^{4/3}$ near $p_c$ (Eq. 1). We can extract a $J_{cN}(\alpha_c)$ scaling relation, by noting from Fig. 3 (or from a Taylor expansion of Eq. (8)) that $p(\alpha_c)$ is approximately linear near $p_c$. Using our rule from Eq. (9), we approximate $p-p_c \sim \alpha_c-\alpha_m$ and find

$$J_{cN}(\alpha_c) \sim (\alpha_c-\alpha_m)^{4/3} \qquad \text{for } \alpha_c >\approx \alpha_m \qquad (10)$$

Fig. 4 shows a comparison of Eq. 10 with data calculated by Rutter et al. [41]. Our formula fits the data well over a surprisingly wide range of $\alpha_c$.

In the more general case of Eq. 4 and $\alpha_0 > 0$, all GBs have nonzero critical currents, and the overall $J_{cN}$ is nonzero for all $\alpha_c > 0$. A general analytical treatment is difficult, but the dependence was studied numerically [46]. It was shown that in this case also, $J_{cN}(\alpha_c)$ has a characteristic scaling behaviour, with 2 approximately linear regimes separated by a kink at $\alpha_c \approx 2\alpha_m$.

## 3.2 Finite-Size Scaling of Critical Currents

From the results of the previous section, we can infer that the finite-size scaling of the critical probability $p_c$ has important consequences for the critical current: Since in the bimodal case $I_{cN}(p < p_c) = 0$, and $p_c$ in systems with finite size $N^D$ scales according to Eq. (3), we expect that the overall critical current density $J_{cN} = I_{cN}/N^{D-1}$ of a conductor must also scale with the conductor cross section $N^{D-1}$. This has been investigated and confirmed by several authors [40-42,44], and is of great practical interest for conductor design. Fig. 5 shows the $J_{cN}$ data calculated by Rutter *et al.* [41] for a 2-dimensional model conductor ($L^{D-1} =$ conductor width $W$). Here we want to point out that these data obey the same scaling law as function of conductor width, if the conductor length is not too large:

$$J_{cN}(\infty) - J_{cN}(W) \sim W^{-1/\nu} , \quad \nu = 4/3 \qquad (11)$$

Hence we have identified a case where a scaling law for a statistical parameter ($p_c$) can be directly applied to an electromagnetic quantity ($J_{cN}$). For the more general case ($\alpha_0 > 0$), no data are available, but we can expect that critical current densities $J_{cN}$ are also dependent on conductor cross section.

## 3.3 Current vs Probability

Next we consider the more general and realistic nonlinear resistor network models described by Eqs (4) and (6), and investigate the most interesting relationship probability $p$ vs. total current $I$. The function $p(I)$ is the number of superconducting GBs divided by total number of GBs, as function of total current. It should be unity at currents smaller than the onset critical current of the conductor, and decrease with larger currents. Fig. 6 shows some typical current and flux distributions generated by the model described earlier [45,46]. Fig. 6a shows the current flowing through the sample from left to right, and Fig. 6b the GB voltage or flux percolation rate. Note that the latter is different from the flux *density* which is measured in magneto-optical measurements (Fig 1). The magnetic flux channels, which consist of moving vortices perpendicular to the paper plane, penetrate the sample from top to bottom. Most flux channels branch out and connect with others in a complex pattern, but all channels reach across the entire sample. This follows within the model from the Kirchhoff voltage rule, and physically from the fact that the moving vortices cannot simply disappear, i.e. the flux channels cannot have 'dead ends' within the sample.

This means also that there can be no isolated resistive GBs; one always finds rows of GBs (channels) spanning the entire conductor width. Hence the probability of the GBs being superconducting is not independent, but correlated as function of the current. The basic assumption of percolation theory is not fulfilled, and we cannot apply its results in a straightforward manner.

In order to restore independence of variables, one might consider using the flux channels as elements of the 'renormalised' statistical ensemble. However, the 'branching-out' of flux channels in all non-trivial cases means that, for a given current, the flux channels partly overlap. Rutter et al. [51] have shown that overlapping channels are not 'independent' as function of total current. That is, the existence of the first branch modifies the onset current of the second one (because it modifies the current distribution in the conductor). Therefore, not even for the flux channels are the individual states (superconducting or resistive) independent.

Several authors [7,37,50] have used a Weibull distribution [52,53] to establish a relation between statistical probability and electromagnetic properties. However, in these analyses the elements of the statistical ensemble were 'short pieces' of conductor, rather than individual GBs. The Weibull distribution, of which the Rayleigh distribution (Eq.s 7, 8) is a special case, also assumes statistical independence of the state of the individual 'elements', and therefore its applicability for individual GBs is doubtful.

In summary, we find a lack of statistical independence in the state of the GBs when a transport current $I > I_{cN}$ is applied and a finite voltage is present. This appears to prevent us from applying percolation theory to describe transport current dissipation in granular superconductors. Moreover, the correlation through the Kirchhoff voltage rule is implicit and non-local, and hence we cannot utilise results of 'standard' theory of correlated percolation. On the other hand, we know that in these correlated systems, scaling laws similar to those of 'random percolation' were found. Furthermore, we have seen that in the limiting case of $\alpha_0 \rightarrow 0$, we reproduce a random resistor network and the associated scaling laws of percolation theory. Hence we can expect to find similar scaling also in our general case.

We have found power-law scaling also in the *I-V* characteristics of nonlinear resistor networks [45,46], as shown in Fig. 7. The inset of the figure shows that in the curved part, the *I-V* curves scale as

$$V \sim (I - I_{cN})^k, \ \ k \approx 1.33. \tag{12}$$

The scaled curves are kinked, because different areas of the sample have different critical currents, as was shown in Ref [46]. The numerical value of the exponent k is close to the random resistor network resistivity exponent $s \approx 1.30$ and the scaling exponent for the coherence length $\nu = 4/3$. Our data were not generated with high numerical accuracy, but we found this scaling consistently in both measured and model samples. We also point out that the values of the voltage scaling exponents probably depend on the shape of the probability distribution function, as was found for related percolative systems [54,55].

We can use this numerical result as a starting point to establish the required relation $p(I)$ in the form of a power-law

$$p - p_c \sim |(I - I_{cN})|^x \qquad (13)$$

We first note that the 'above-percolation' domain ($p > p_c$) refers to the 'below-critical current' situation ($I < I_{cN}$) and vice versa. Secondly, as a consequence of the Kirchhoff rules, the critical probability $p_c$ is always unity for a current-induced transition. If at least one GB becomes resistive ($p < 1$), a whole channel must become resistive simultaneously. This channel cuts through the infinite/spanning superconducting 'cluster', and spans the whole sample with a normal conducting 'cluster' perpendicular to the current direction.

From (12) we find

$$\rho = dV/dI \sim (I - I_{cN})^{k-1}. \qquad (14)$$

Comparison of Eqs (2), (13) and (14) gives x = (k–1)/s. For the 2-*D* case (s $\cong$ 1.3) we find x $\cong$ ¼.

In order to test the validity of this result, we can check it against data from our nonlinear resistor network model. Fig. 8 shows $p(I)$ data (as defined above) from the sample described in Figures 6 and 7, for different values of $\alpha_c$. The first observation is that the function $p(I)$ is discontinuous. This is not a numerical error, but a consequence of the finite sample size in combination with the Kirchhoff rules. The first jump from $p = 1$ to $p < 1$ appears at the onset critical current $I_{cN}$, representing the appearance of the first flux channel. Even at higher currents, $p(I)$ is not smooth, but has several shoulders and jumps, which correspond to the linear segments and kinks in the scaled *I-V* curves (inset to Fig. 7). Therefore, an approximation with a continuous distribution function such as Eq. (13) cannot be expected to be accurate. We have fitted Eq. (13) with x = ¼ to the onset regime (first jump and shoulder) of some of the data ($\alpha_c = 2, 4$). The fits are shown as dashed lines in Fig. 8, and show qualitative agreement. This result needs to be confirmed on larger systems where finite-size effects are less dominant.

## 4 Experimental Current-Voltage Data

The current-voltage scaling described in the previous section (Eq. 12) can also be found in experimental data measured on YBCO coated conductors and Bi2223 tapes, however with different exponents *k*.

Fig. 9 shows *I-V* curves from a coated conductor, measured at 77 K in various magnetic fields. The inset shows the data plotted as $V^{1/k}$, k = 2.5, against *I*, and in this plot they appear as straight lines. We investigated a number of different samples in the same way, and found that they all scaled according to Eq. 12. For each sample the exponent *k* was constant as function of magnetic field, but the exponents varied between 1.5 and 4 for different samples. None of the samples showed the lower exponent k ≈ 1.3 found in the resistor network simulations. This difference is possibly related to the fact that the simulation is based on a simple model for the GB *I-V* with only one kink, while *I-V* characteristics of real GBs more often show several kinks [27].

In Bi2223 tapes we also found percolation scaling of *I-V* curves (Eq. 12), but only in zero field data. Fig 10 shows current-voltage curves measured in a pressurised liquid nitrogen vessel [56,57], in self field at different temperatures. The inset plot shows that the curves scale according to Eq. (12) but with $k = 4$. Note that at high voltages the curves deviate from the scaling. This is caused by the fact that with higher voltage, progessively more current flows through the silver sheath of the tape.

If one has a high density of *I-V* data points, as we have in the case of the Bi2223 tape, the *I-V* scaling can also be investigated by calculating the logarithmic derivative of the *I-V* [57,58]. From Eq. (12) follows

$$\mathrm{d}(\ln V)/\mathrm{d}(\ln I)_{\mathrm{perc}} = k/(1 - I_c/I). \tag{15}$$

Hence, if we plot this quantity as function of $I/I_c$, a set of *I-V* curves with constant $k$ should scale as a single, hyperbolic curve. This is what we find for the zero-field *I-V* data, as shown in Fig. 11. The exponent extracted from fitting Eq. (15) to these data (solid line in Fig. 11) is $k = 4$. Note that by calculating the logarithmic derivative from the *I-V* curve, we have reduced the number of fit parameters, and thus can extract the exponent with higher confidence. Comparing this result with percolation theory for 3-*D* networks ($s_{3D} \cong 0.73$) and applying the same scaling analysis as above (Eq.s 13 and 14), we find for the current – probability scaling exponent

$$x_{3D} = (k-1)/s_{3D} \cong 4 \tag{16}$$

This result is purely speculative until it can be tested in a 3-*D* computer simulation.

Fig. 11 also includes logarithmic derivatives of *I-V* data measured at 50 mT. These data also scale as one curve, which however is not hyperbolic but approximately constant. This results suggest a power-law *I-V* dependence $V \sim I^n$ which can be attributed to flux creep [59]. We conclude that only in zero or small magnetic field the current transfer in Bi2223 tapes can be described as a percolative process, while in magnetic field the dissipation is controlled by flux pinning and de-pinning. This interpretation conforms with irradiation experiments by Tönies *et al.* [35]. They also concluded that in zero and small field grain boundaries dominate critical currents in Bi2223 conductors, and in higher fields intra-grain pinning is the dominating factor.

## 5. Summary

In this paper we have attempted to link the concepts of percolation theory to current transfer and dissipation in polycrystalline superconductors. Since percolation theory deals mainly with statistical concepts, and in experiments electromagnetic quantities are measured, computer simulations provide a vital tool, because in these both statistical and electromagnetic quantities are accessible.

The network of grains and grain boundaries can be considered as a statistical ensemble, with the GB state (superconducting or resistive) as the statistical variable. In most percolation treatments of superconductivity, the GB state is controlled by random processes, e.g. temperature fluctuations. If however the superconducting state is determined by the overall transport current, statistical independence of the GB states is lost. The applicability of percolation theory is then questionable. On the other hand, we have shown that also in this case, we can identify scaling behaviour that is characteristic for percolative processes.

The immediate tasks for future work are to confirm our simulation results on larger systems, and to find a theoretical foundation for them. The latter can possibly be found by expanding the concepts of 'standard' theory for correlated percolation to the type of correlation described by the Kirchhoff rules.

In practical terms it is clear that percolation theory can be used to predict essential properties of polycrystalline superconductors (e.g. Equations (10), (11) and (14)), which is important for optimisation of practical conductor design.

We thank Christian Jooss (Goettingen University) for providing Fig. 1 and Noel Rutter (now at Oak Ridge National Lab) for data shown in Figures 4 and 5. Susi Toenies, Yanfeng Sun and Alex Vostner (Atomic Institute Vienna) supplied coated conductor data (Fig. 9). Bi2223 tapes were generously provided by NST A/S, Brondby, Denmark. BZ wishes to thank Prof. H.W. Weber and other members of the Atomic Institute in Vienna for a very pleasant and fruitful visit, during which many of the ideas in this paper were developed.
This work was carried out as part of the European Union TMR Network SUPERCURRENT.

## References

1. A. Davidson and M. Tinkham, Phys. Rev. B **13**, 3261 (1976)
2. G. Deutscher, O. Entin-Wohlman, S. Fishman and Y. Shapira; Phys Rev B **21**, 5041 (1980)
3. E. Bruneel, A.J. Ramirez-Cuesta, I. van Driessche and S. Hoste, Phys. Rev. B **61**, 9176 (2000)
4. K.V. Mitsen and O.M. Ivanenko, Physica C **341-8**, 1849 (2000)
5. H.A. Knudsen and A. Hansen, Phys. Rev. B **61**, 11336 (2000)
6. N. Markovic, C. Christiansen, A.M. Mack, W.H. Huber, and A.M. Goldman, Phys. Rev. **B** 60, 4320 (1999)
7. K. Yamafuji and T. Kiss, Physica C **258**, 197 (1996)
8. M. Ziese, Phys. Rev. B **55**, 8106 (1996)
9. M. Prester, Phys. Rev. B **54**, 606 (1996)
10. M. Prester, Supercond. Sci. Techn. **11**, 333 (1998)
11. M. Sahimi, *Applications of Percolation Theory* (Taylor & Francis, London, 1994)
12. D. Stauffer and A. Aharony, *Introduction to Percolation Theory*; 2nd edn (Taylor & Francis, London, 1992)
13. J.P. Clerc, G. Giraud, J.M. Laugier and J.M. Lucks, Adv. Phys. **39**, 191 (1990)
14. S. Kirckpatrick, Rev. Mod. Phys. **45**, 574 (1973)
15. O. Stennul and H.K. Janssen, Phys. Rev. E **65**, 036124 (2002)
16. A. Coniglio, C.R. Nappi, F, Peruggi and L. Russo, J. Phys. A **10**, 205 (1977)
17. D.W. Heermann and D. Stauffer, Z. Phys. B **44**, 339 (1981)
18. A.L. Stella and C. Vanderzande, Phys. Rev. Lett. **62**, 1067 (1989)
19. D. Dimos, P. Chaudari and J. Mannhart, Phys. Rev. **B** 41, 4038 (1990)
20. S.E. Babcock and J.L. Vargas, Annu. Rev. Mater. Sci. **25**, 193 (1995)
21. N.F. Heinig, R.D. Redwing, J.E. Nordman and D.C. Larbalestier, Phys. Rev. B **60**, 1409 (1999)
22. J.G. Wen, T. Takagi and N. Koshizuka, Supercond. Sci. Techn. **13**, 820 (2000)
23. J. Hänisch, A. Attenberger, B. Holzapfel, and L. Schultz, Phys. Rev. B **65**, 052507 (2002)
24. K.E. Gray et al., Physica C **341-348**, 1397 (2000)

25. G. Hammerl et al., IEEE Trans. Appl. Supercond. **11**, 2830 (2001)
26. A. Diaz, L. Mechin, P. Berghuis and J.E. Evetts, Phys. Rev. Lett. **80**, 3855 (1998)
27. M.J. Hogg, F. Kahlmann, E.J. Tarte, Z.H. Barber and J.E. Evetts, Appl. Phys. Lett. **78,** 1433 (2001)
28. J.E. Evetts, M.J. Hogg, B.A. Glowacki, N.A. Rutter and V.N.Tsaneva, Supercond. Sci. Techn. **12**, 1050 (1999)
29. A. Goyal, E.D. Specht, D.K. Christen, D.M. Kroeger, A. Pashitski, A. Polyanskii and D.C. Larbalestier, J. of Materials, October 1996, 24
30. D.C. Larbalestier, A. Gurevich, D.M. Feldmannn and A. Polyanskii, Nature 414, 368 (2001)
31. D.M. Feldmann et al., IEEE Trans. Appl. Supercond. **11**, 3772 (2001)
32. C. Jooss et al., MRS Proceedings **659** (2001) II 7.1
33. P. Vase, R. Flukiger, M. Leghissa and B.A. Glowacki, Supercond. Sci. Techn. **13**, R71 (2000)
34. A. Polyanskii, D.M. Feldmann, S. Patnaik, J. Jiang, X. Cai, D. Larbalestier, K. DeMoranville, D. Yu, R. Parella, IEEE Trans. Appl. Supercond. **11**, 3269 (2001)
35. S. Tonies, H.W. Weber, Y.C. Guo, S.X. Dou, R. Sawh and R. Weinstein, Appl. Phys. Lett. **78,** 3851 (2001)
36. E.D. Specht, A. Goyal and D.M. Kroeger, Phys. Rev. B **53**, 3585 (1996)
37. K. Osamura, K. Ogawa, T. Thamizavel and A. Sakai, Physica C **335**, 65 (2000)
38. R. Haslinger and R. Joynt, Phys. Rev. B **61**, 4206 (2000)
39. H.A. Knudsen and A. Hansen, Phys. Rev. B **61**, 11336 (2000)
40. E.D. Specht, A. Goyal, and D.M. Kroeger, Supercond. Sci. Techn. **13**, 592 (2000)
41. N.A. Rutter, B.A. Glowacki and J.E. Evetts, Supercond. Sci. Techn. **13**, L25 (2000)
42. R. Mulet, O. Diaz, and E. Altshuler, Supercond. Sci. Techn. **10**, 758 (1997)
43. B. Holzapfel et al., IEEE Trans. Appl. Supercond. **11**, 3872 (2001)
44. Y. Nakamura, T. Izumi and Y. Shiohara, Physica C **371**, 275 (2002)
45. B. Zeimetz, N.A. Rutter, B.A. Glowacki and J.E. Evetts, Supercond. Sci. Techn. **14,** 672 (2001)
46. B. Zeimetz, B.A. Glowacki and J.E. Evetts, Physica C (in press)
47. D.M. Feldman et al., Appl. Phys. Lett. **77**, 2906 (2000)
48. P.A. Jensen and J.W. Barnes, *Network Flow Programming* (John Wiley & Sons, New York, 1980)
49. D.T. Philips and A. Garcia-Diaz, *Fundamentals of Network Analysis* (Prentice-Hall Inc, Englewood Cliffs, NY, 1981)
50. N.A. Rutter, Ph.D thesis, Cambridge University (2001), available also at ftp://dmg-stairs.msm.cam.ac.uk/pub/theses
51. N.A. Rutter and B.A. Glowacki, Supercond. Sci. Techn. **14**, 680 (2001)
52. W. Weibull, J. Appl. Mech. **18**, 293 (1951)
53. L.J. Bain, *Statistical Analysis of Reliability and Life-Testing Models* (Marcel Dekker, New York, 1978)
54. S. Roux and H.J. Hermann, Europhys. Lett. **4**, 1227 (1987)
55. M. Sahimi and J.D. Goddard, Phys. Rev. B **33**, 7848 (1986)
56. B.A. Glowacki, E.A. Robinson and S.P. Ashworth, Cryogenics **37**, 173 (1997)

57. B. Zeimetz, B.A. Glowacki, Y.S. Cheng, A. Kursumovic, E. Mendoza, X. Obradors, T. Puig, S.X. Dou and J.E. Evetts, Inst. Phys. Conf. Ser. **167**, 1033 (2000)
58. P. Berghuis, R. Herzog, R.E. Somekh, J.E. Evetts, R.A. Doyle, F. Baudemacher, and A.M. Campbell, Physica C **256,** 13 (1996)
59. E. Zeldov, N.M. Amer, G. Koren, A. Gupta, M.N. McElfresh, R.J. Gambino, Appl. Phys. Lett. **56**, 680 (1989)

## Figure Captions

Fig. 1: Magneto-optical image of a YBCO coated conductor, taken at 5K in a perpendicular field of 80 mT; the local magnetic field is shown as a grayscale (courtesy of Christian Jooss)

Fig. 2: Percolation maps of a polycrystalline coated conductor at different cutoff angles $\alpha_c$ as indicated; GB angles $\alpha > \alpha_c$ are shown as solid lines, the largest cluster of connected grains is shown in gray, the arrows in (c) and (d) indicate the location of the 'limiting path', i.e. the bottleneck for supercurrent flow

Fig. 3: Main plot: cumulative probability $p(\alpha > \alpha_c)$ of a Rayleigh distribution (see text, Eq 8), with percolation thresholds $p_c$ for hexagonal and square 2-*D* lattices indicated; Inset: GB angle frequency; line: probability distribution function $dp/d\alpha$ (Eq. 7); histogram data from sample shown in Fig. 2

Fig. 4: Data points: normalised critical current, expressed as limiting cross section for supercurrent, as function of threshold angle $\alpha_c$, for a hexagonal model conductor with $8 \times 9$ grains, from [37], dotted line: dependence $I_c \sim (\alpha_c - \alpha_m)^{4/3}$ (Eq. 10)

Fig. 5: Onset critical current against conductor width for different conductor lengths in a hexagonal model conductor, in a bimodal calculation, data from NA Rutter *et al.* [37]. Solid lines: fit to finite-size scaling law (Eq. 11) $J_{cN}(\infty) - J_N(W) \sim W^{-1/\nu}$, $\nu = 4/3$

Fig. 6: Calculated current and flux distribution in a grain network. The current flows through the sample from left to right, a) grain current distribution (*not* normalised), shown as a grey scale; b) percolation rate of magnetic flux (= GB voltage), shown as white lines with line thickness proportional to flux

Fig. 7: Main plot: calculated current-voltage curves of the sample shown in Fig. 6, for different values for $\alpha_c$ as indicated; Inset: the same data of curved region, plotted as $V^{3/4}$ against $I$, showing piecewise linear segments and kinks (see text)

Fig. 8: Number of superconducting GBs divided by total number of GBs, against total current, for different values of $\alpha_c$ as indicated; solid lines: dependence $p \sim 1 - (I - I_{cN})^{1/4}$ fitted to data near onset; dotted lines are guides to the eye

Fig. 9: Main plot: current-voltage curves of a coated conductor sample, measured at 77 K in various magnetic fields as indicated; Inset: same data, plotted as $V^{0.4}$ against current; demonstrating scaling behaviour $V \sim (I - I_{cN})^{2.5}$

Fig. 10: Main plot: current-voltage curves of a Bi2223 tape with 37 filaments, measured in a pressurised liquid nitrogen vessel, in self field at various temperatures as indicated; Inset: same data, plotted as $V^{1/4}$ against current

Fig. 11: Logarithmic derivative $d(\ln V)/d(\ln I)$ of the Bi2223 tape *I-V* data shown in Fig. 10 (open symbols, self field), and of *I-V* data measured in 50 mT (solid symbols); at various temperatures as indicated

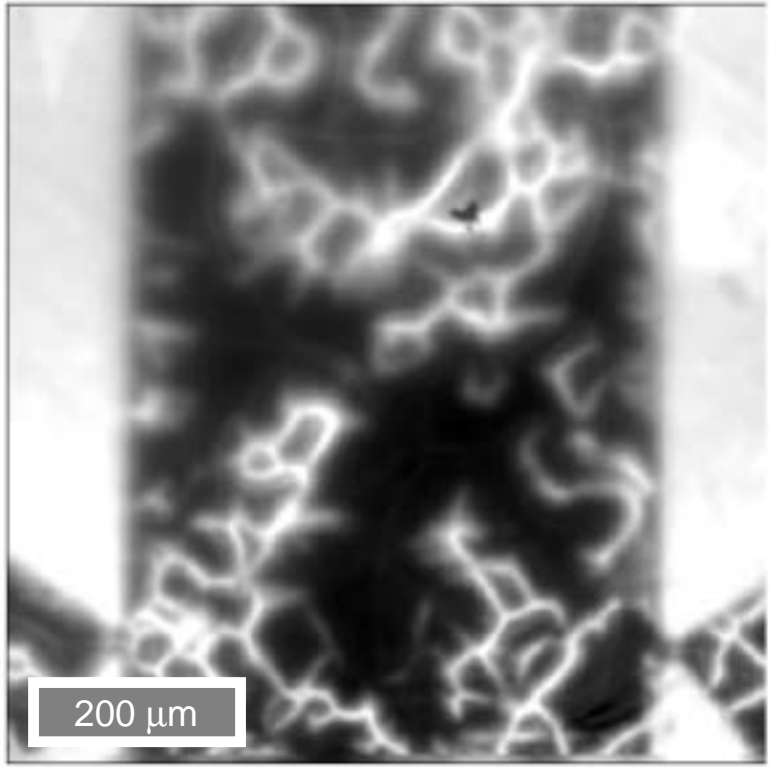

200 µm

Figure 1

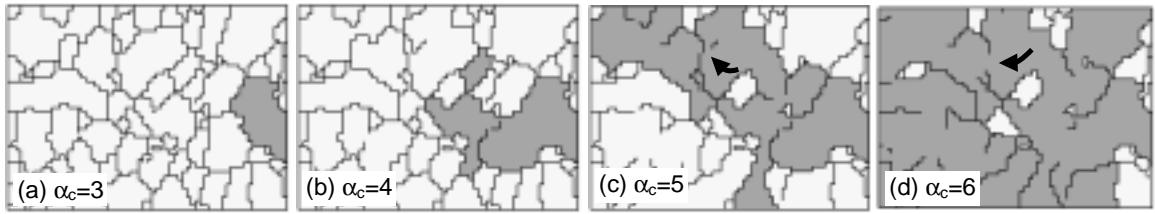

Figure 2

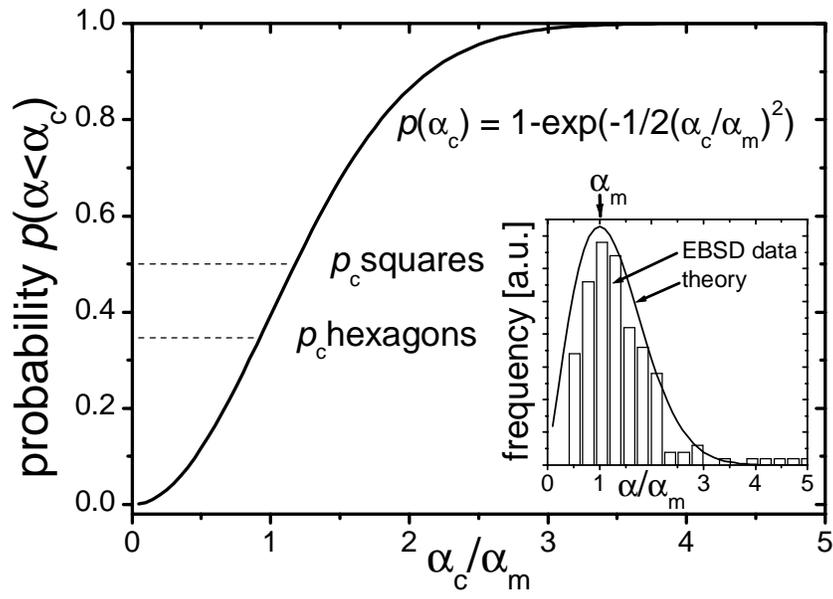

Figure 3

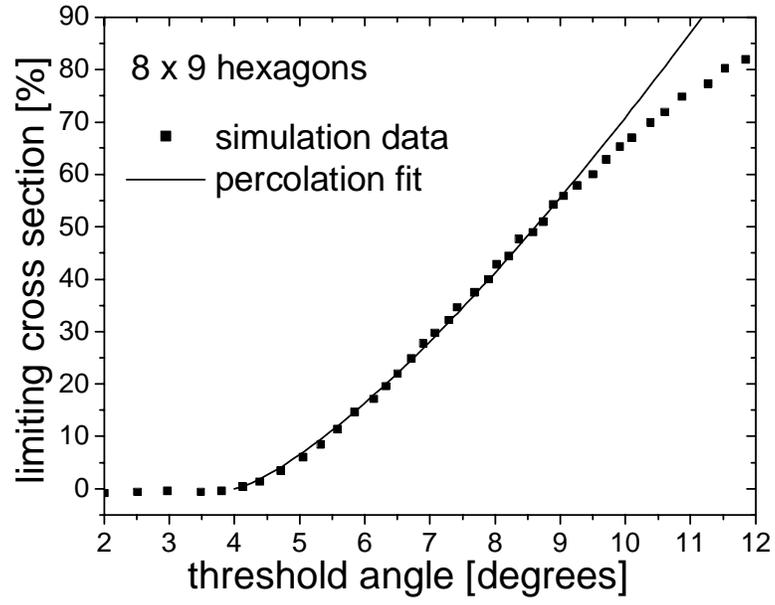

Figure 4

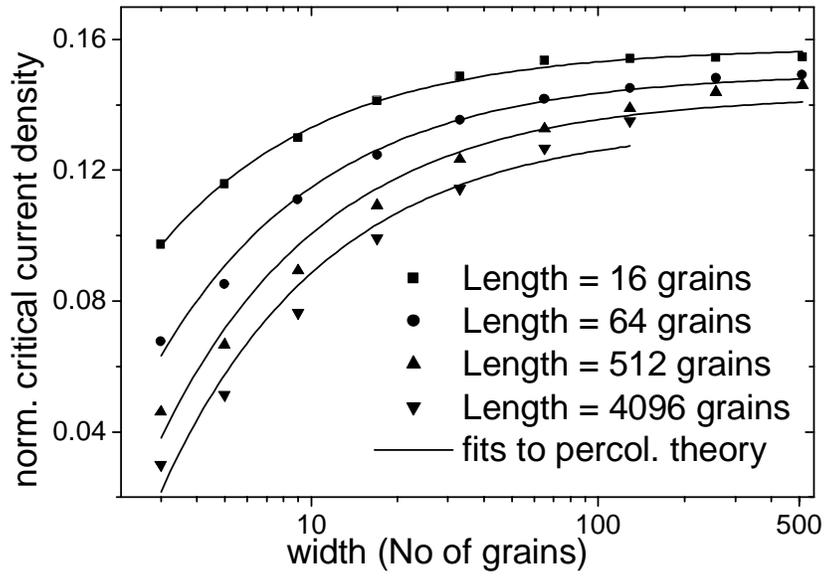

Figure 5

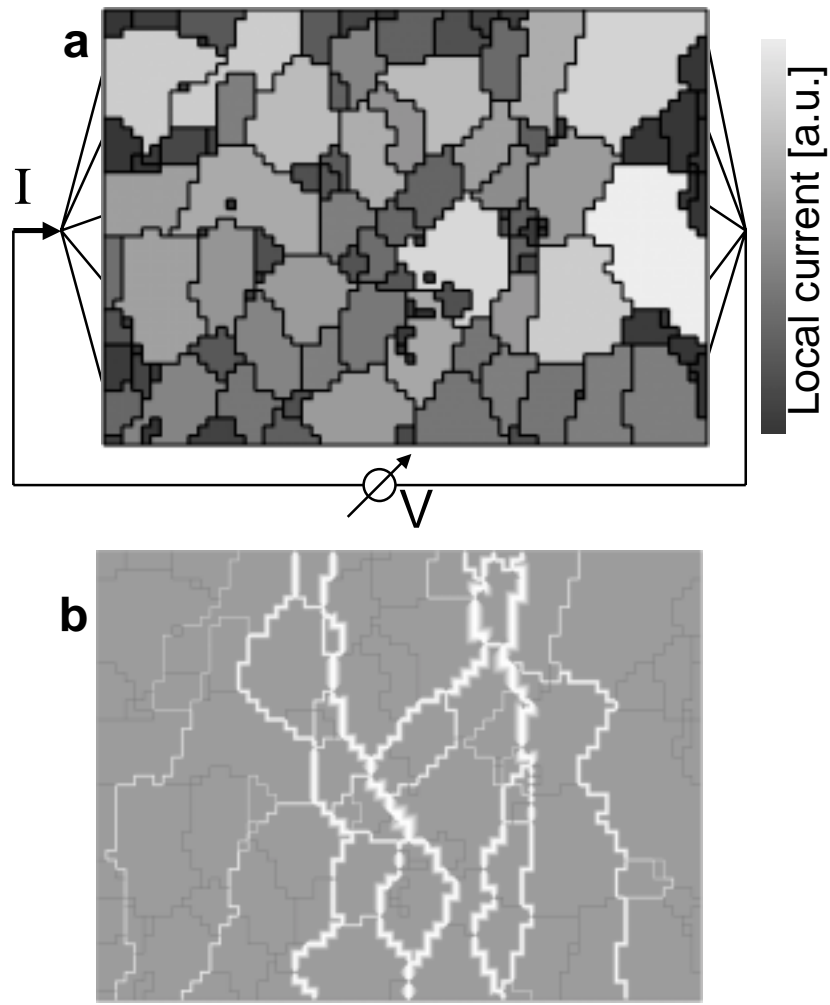

Figure 6

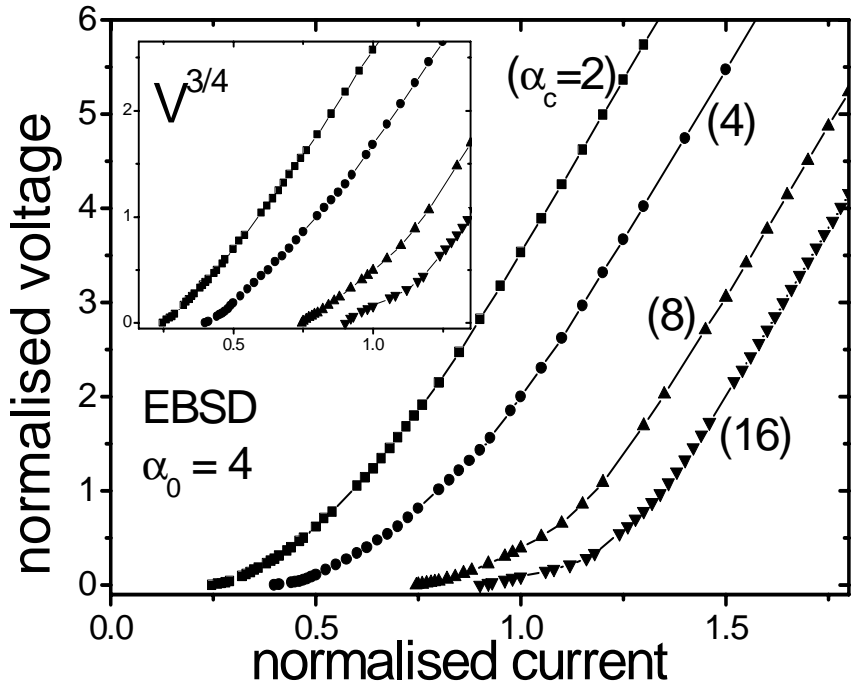

Figure 7

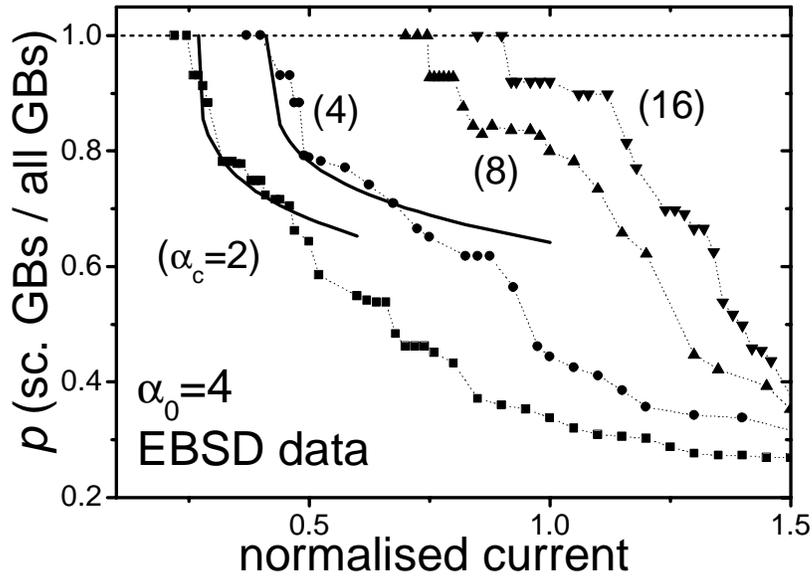

Figure 8

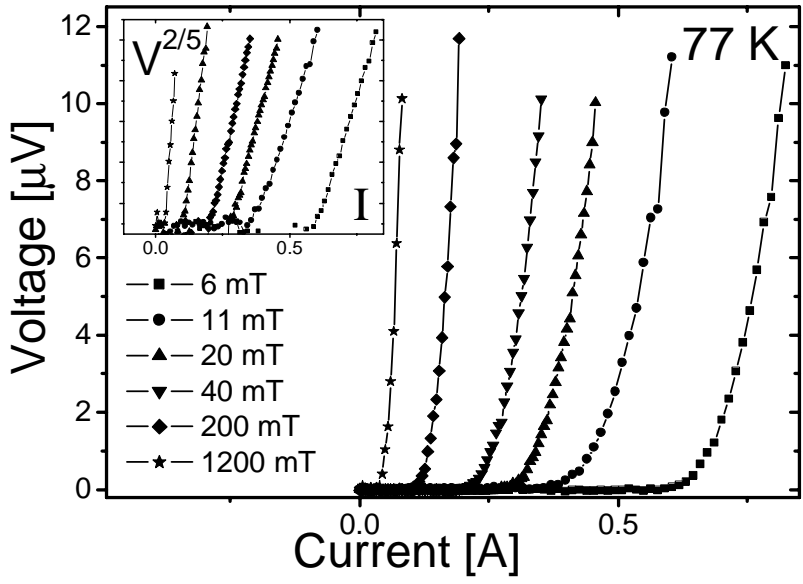

Figure 9

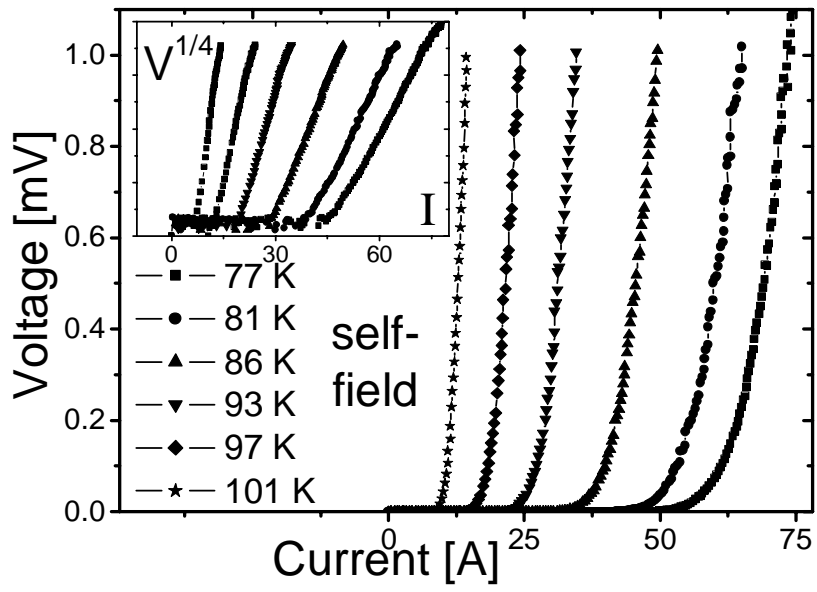

Figure 10

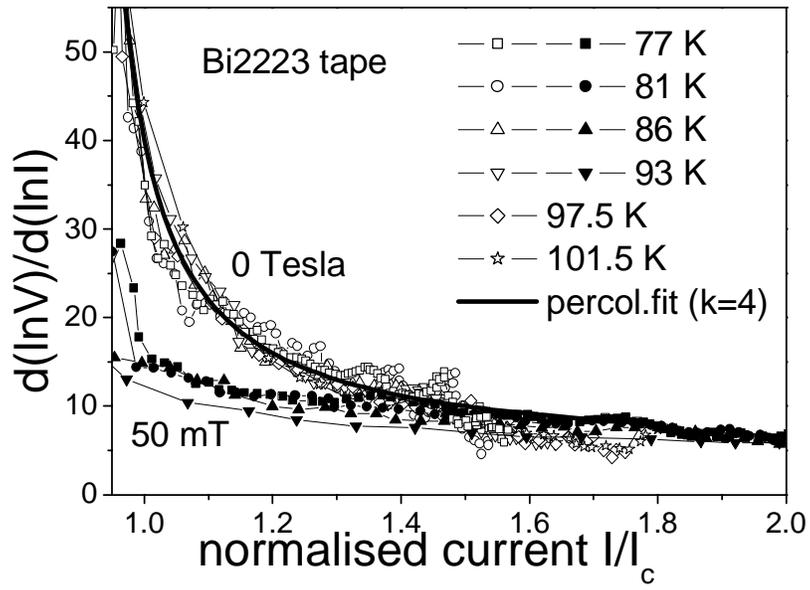

Figure 11